\documentclass{article}
\usepackage{RR}
\RRNo{6270}
\usepackage{epsfig} %% add this line and the next only if you have pictures
\usepackage{graphics} %% pictures should be in esp format
\usepackage[T1]{fontenc}
\usepackage[latin1]{inputenc}
\usepackage{amsfonts}
\usepackage{amsmath}
\usepackage{amssymb}

\newcommand \dps{\displaystyle }

\RRdate{August, 2007}
\RRauthor{% les auteurs
R. Alicandro$^1$, M. Cicalese$^2$, A. Gloria$^3$ \thanks{$^1$ DAEIMI, Universit\`a di Cassino,
via Di Biasio 43, 03043 Cassino (FR), Italy ;
$^2$ Dipartimento di Matematica e Applicazioni `R.~Caccioppoli',
Universit\`a di Napoli,
via Cintia, 80126 Napoli, Italy ;
$^3$ CERMICS - ENPC \& INRIA Paris-Rocquencourt,
6 et 8 avenue Blaise Pascal - Champs sur Marne, France ;
Contact: alicandr@unicas.it, cicalese@unina.it, antoine.gloria@inria.fr}
}
\authorhead{R. Alicandro, M. Cicalese, A. Gloria}
\RRtitle{Propri\'et\'es de convergence du mod\`ele discret de B\"ol-Reese pour le caoutchouc}
\RRetitle{Convergence analysis of the B\"ol-Reese discrete model for rubber}

\titlehead{Convergence of a discrete model for rubber}
\RRresume{Dans~\cite{Bol-Reese-03-1}, B\"ol et Reese ont introduit un mod\`ele discret de r\'eseau de polym\`eres par \'el\'ements finis. Dans~\cite{Bol-Reese-05c,Bol-Reese-05}, les auteurs d\'etaillent leur m\'ethode et comparent les r\'esultats num\'eriques obtenus \`a des exp\'eriences r\'eelles. Un des param\`etres cl\'es de leur mod\`ele est la taille du pas $h$ du maillage, qui est petit en pratique. 
L'objectif de ce travail est d'\'etudier le comportement asymptotique (ainsi que la convergence de la m\'ethode des \'el\'ements finis) lorsque $h$ tend vers z\'ero. En particulier, nous \'etudions les propri\'et\'es satisfaites par le mod\`ele \`a la limite, en fonction, des propri\'et\'es du maillage. 
}
\RRabstract{In~\cite{Bol-Reese-03-1}, B\"ol and Reese have introduced a discrete model for polymer networks by means of a finite element modeling. In~\cite{Bol-Reese-05c,Bol-Reese-05}, they present details on the method and a comparison with real experiments. A key parameter of their model is the size $h$ of the finite element mesh, that is meant to be small in practice. The aim of the present work is to study the asymptotic behaviour (and the convergence of the finite element method) when the meshsize goes to zero. In particular, we address the properties satisfied by the model at the limit, depending on the properties of the mesh. }
\RRmotcle{\'elasticit\'e non lin\'eaire, caoutchouc, syst\`eme discret, m\'ethode des \'el\'ements finis}
\RRkeyword{nonlinear elasticity, rubber, discrete system, finite element method}
 \RRprojet{MICMAC}
 \RRtheme{\THNum} % cas d'un seul theme
\URRocq % pour ceux qui sont au centre de la France
\begin{document}

\makeRR   % cas d'un rapport de recherche

\newcommand{\calT}{\mathcal{T}}

Phenomenological constitutive laws for rubber-like materials often involve parameters which lack of physical motivation. In addition, they are usually difficult to fit in practice. Much attention has been paid in the recent years to microscopically-based models. This kind of models aims at pointing out the microscopic features that govern the macroscopic behavior of the material, such as the geometry of the underlying polymer network. Such a model has been introduced by B\"ol and Reese in~\cite{Bol-Reese-03-1}. In addition to the physical parameters provided by physics at the scale of the polymer chains (such as the free energy of a chain), this model involves two geometric parameters: a typical distance $h>0$ (which is meant to be small) and a tetraedral mesh.

\medskip

The aim of the present work is to study the asymptotic properties of the model when $h$ goes to zero for different assumptions on the mesh.

\medskip

This note is organized as follows:

\tableofcontents

\newpage

\section{The B\"oL-Reese model}

\subsection{The geometry}

For the sake of simplicity, let $\Omega$ denote the unit cube of $\mathbb{R}^3$ and $\calT_h$ be a tetraedral mesh of $\Omega$. The typical distance we associate with the mesh is defined as follows:
$$h=\sqrt[3]{\frac{1}{N_{\mbox{el}}}},$$
where $N_{\mbox{el}}$ denotes the number of elements (tetraedras) of $\calT_h$.

From a modeling point of view, each edge of the mesh represents an elastic spring with an energy related to the underlying polymer network.
 
Let $d$ be a displacement field of the mesh 
defined at each vertex $x_i$ of $\calT_h$ by $d_i$. One can associate with $d$ a continuous displacement field on $\Omega$ as follows: $v_h$ is the unique piecewise affine function on $\calT_h$ such that for each vertex $x_i$ of $\calT_h$, $v_h(x_i)=x_i+d_i$.

The energy of the system deformed by $v_h$ is then given by
\begin{equation}\label{eq:ener-h}
E_h(v_h)=\sum_{T\in \calT_h}h^3 \sum_{(x_i,x_j)\in T}f_{ij}^T W_{ij}^T \left(\frac{|v_h(x_i)-v_h(x_j)|}{|x_i-x_j|} \right)+\int_\Omega W_{vol}(\nabla v_h),
\end{equation}
where $T$ denotes elements of $\calT_h$, $(x_i,x_j)$ are distinct pairs of vertices in the tetraedra $T$, $W_{ij}^T$ is a pair potential, $f_{ij}^T$ is a scaling factor and $W_{vol}$ is an energy density accounting for volume changes of the deformed system.

Let us now describe the pair-potential energy.

\subsection{The energy}

The energy of the elastic springs is related to the energy of a polymer chain as follows:
\begin{equation}\label{eq:ener-polch}
W_{ij}^T(r)=\frac{k}{\beta}n_{ij}^T\left( \frac{r}{\sqrt{n_{ij}^T}}\mathcal{L}^{-1}\left(\frac{r}{\sqrt{n_{ij}^T}}\right)+\ln \frac{\mathcal{L}^{-1}\left(\frac{r}{\sqrt{n_{ij}^T}}\right)}{\sinh\mathcal{L}^{-1}\left(\frac{r}{\sqrt{n_{ij}^T}}\right)} \right) - \frac{c}{\beta}, 
\end{equation}
where $\beta$ is the inverse of the absolute temperature, $\mathcal{L}^{-1}$ is the inverse of the Langevin function, $k$ and $c$ are constants and $n_{ij}^T$ is a typical number of segments of a polymer chain. $W_{ij}^T$ can be interpreted as the free energy of a polymer chain made of $n_{ij}^T$ segments at the length $\sqrt{n_{ij}^T}l$, where $l$ denotes the length of a segment.

The factor $f_{ij}^T$ is a measure of the number of polymer chains per unit of volume. We will typically consider it as constant on $\Omega$.

Finally, the energy accounting for volume changes is given by
\begin{equation}\label{eq:ener-vol}
W_{vol}(\xi)=\frac{K}{4}(J^2-1-\ln J), 
\end{equation}
where $K>0$, $\xi$ is a deformation gradient and $J=\det(\xi)$.
The physical origin of this contribution is the Van der Waals forces, which prevent atoms from being too close to one another. The scale of this forces is far larger than the scale of the polymer network, which explains why this contribution is already coarse-grained and somewhat uncorrelated with the description of the network.

In the following section, we study the convergence when $h\to 0$ of the sequence of minimization problems
$$\inf \left\{ E_h(v_h),v_h \in V_h+BC \right\}, $$
where $V_h$ denotes the space of piecewise affine functions on $\calT_h$ and $BC$ stands for the boundary conditions (let say mixed imposed displacement and free traction for instance).
In particular, we address both the convergence of the infimum of the energy and the convergence of the minimizers (the deformation field) when $h\to 0$. This convergence analysis is performed using $\Gamma$-convergence (see \cite{Braides-02,Braides-06} e.g.).

\section{Convergence analysis}

\subsection{Assumptions on the energies}

For technical reasons, in what follows, we will assume that:
\begin{itemize}
\item $W_{ij}^T$ satisfies the following standard growth condition of order $p>1$ for all $T$ and $ij$: there exist $C\geq c>0$ such that for all $r$,
\begin{equation}\label{eq:gc}
c|r|^p-1 \leq W_{ij}^T(r) \leq C(|r|^p+1),
\end{equation}
\item $W_{vol}$ satisfies~(\ref{eq:gc}) from above.
\end{itemize}

In particular, one cannot directly deal with the inverse of the Langevin function. To satisfy the growth condition~(\ref{eq:gc}), one may use a truncated series expansion of $\mathcal{L}^{-1}$, as it is done in~\cite{Bol-Reese-03-1,Bol-Reese-05c,Bol-Reese-05}:
$$\mathcal{L}^{-1}(\rho)=3\rho+\frac{9}{5}\rho^3+\frac{297}{175}\rho^5+\frac{1539}{875}\rho^7+O(\rho^9).$$
Replacing $\mathcal{L}^{-1}$ by its development up to order 7 in~(\ref{eq:ener-polch}), we obtain a free energy density for a polymer chain which satisfies~(\ref{eq:gc}) for $p=8$. 

\medskip

The energy density $W_{vol}$ given by~(\ref{eq:ener-vol}) does not satisfy~(\ref{eq:gc}) since $\lim_{\det(\xi) \to 0} W_{vol}(\xi)=+\infty$ while $\det(\xi)\to 0$ does not imply $|\xi|\to \infty$. One possible solution consists in taking a cut-off, which amounts to relaxing the constraint of incompressibility.
A typical cut-off reads as follows:
\begin{equation}
W_{vol}^\eta(\xi)=\left\{ \begin{array}{ll}
W_{vol}(\xi) & \mbox{ if }\det(\xi)>\eta \\
\frac{K}{4}(\eta^2-1-\ln \eta) & \mbox{ otherwise.}
                          \end{array}
\right.
\end{equation}
For $\eta>0$ fixed, $W_{vol}^\eta$ satisfies~(\ref{eq:gc}) for $C_\eta$ large enough. To recover the incompressibility behavior, one can first perform the convergence analysis on the discrete to continuum process for $\eta >0$ and then let $\eta$ go to zero at the continuum level. We refer the reader to \cite{Gloria-07,Alicandro-Cicalese-Gloria-07b} for technical details. 

\medskip

We are now in position to state our convergence results.

\subsection{The periodic case}

\begin{figure}[tbh]
  \centering
  \includegraphics[height=3cm]{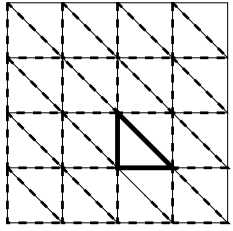}
  \caption{Two-dimensional periodic mesh, the unit cell is in bold \label{fig:mesh-2D}}
\end{figure}

In this paragraph, we address the convergence of the B\"ol-Reese model in the periodic case. To this end, let us assume that the edges of $\calT_h$ are obtained by the periodic replication of a unit cell of edges (up to border effects on the boundary), as illustrated in two dimensions on Figure~\ref{fig:mesh-2D}.

Let us further assume that the energy of the elastic springs does not depend of the unit cell and on the mesh lengthscale $h$. As a particular case, one can take $n_{ij}^T$ to be constant on $\calT_h$, as it is done in~\cite{Bol-Reese-03-1,Bol-Reese-05c,Bol-Reese-05}. Then, \cite[Theorem~4.1]{Alicandro-Cicalese-04} shows there exists an homogeneous quasiconvex energy density $W_{hom}$, which is frame-invariant and satisfies a growth condition of order $p$, such that:
\begin{itemize}
\item $\dps{\lim_{h\to 0} \inf \left\{ E_h(v_h),v_h \in V_h+BC \right\} = \inf \left\{ E_{hom}(v),v \in V+BC \right\}}$, where $V$ is the Sobolev space $W^{1,p}(\Omega)$, and 
$$E_{hom}(v)=\int_\Omega W_{hom}(\nabla v); $$
\item $W_{hom}$ is given by the asymptotic discrete homogenization formula
$$W_{hom}(\xi)=\lim_{h \to 0} \inf \{ E_h(v_h), v_h \in V_h, v(x_i)=\xi\cdot x_i \mbox{ if }d(x_i,\partial \Omega)\leq 2h \}.$$
\end{itemize}
In addition, minimizers $u_h$ of $E_h$ on $V_h+BC$ weakly converge in $W^{1,p}(\Omega)$ to minimizers $u_{hom}$ of $E_{hom}$ on $V+BC$.
This proves in particular the convergence of the solutions of the finite element modelling of rubber introduced by B\"ol and Reese.

\medskip
 
In the periodic case, the energy density at the limit is frame-invariant. However it is not isotropic in general. To ensure the isotropy, one may use a stochastic framework.
The end of this paragraph is dedicated to the study of a simple example for which the energy density at the limit is not isotropic.

Let us consider simple linear springs by setting
$$W_{ij}^T(v_h)=K\left(\frac{|v_h(x_i)-v_h(x_j)|}{|x_i-x_j|}\right)^2 \quad \mbox{and} \quad W_{vol}=0$$
on the mesh sketched on Figure~\ref{fig:mesh-2D}.
% \begin{figure}[tbh]
%   \centering
%   \includegraphics[height=3cm]{mesh-2D-3.eps}
%   \caption{Two-dimensional orthotropic mesh (checkerboard) \label{fig:mesh-2D-2}}
% \end{figure}
Then, recalling~\cite[Remark~5.2]{Alicandro-Cicalese-04}, the homogenized energy density is quadratic and we end up with a convex minimization problem on one single periodic cell. The infimum is trivial and shows that the homogenized energy is not isotropic. It is actually enough to see that the material is stiffer in the direction $e_2-e_1$ than in the direction $e_1+e_2$ (where $e_1$ and $e_2$ denote the canonical basis of $\mathbb{R}^2$).

\subsection{The stochastic case}

Let us quickly recall the concept of admissible stochastic network on a probability space $(\Xi,\mathcal{F},P)$ used in~\cite{Alicandro-Cicalese-Gloria-07a,Alicandro-Cicalese-Gloria-07b}, which is a particular case of the stochastic networks introduced by Blanc, Le Bris and Lions in \cite{BLL-06,BLL-07}.

\medskip

Let $\Lambda=\{ y_i \}_{i\in \mathbb{Z}^d}\in \left( \mathbb{R}^d
\right)^{\mathbb{Z}^d}$ be a set of points. We say that $\Lambda$ is
an admissible set of points if it satisfies the two following
conditions:
\begin{itemize}
\item[i.] there exists $R>0$ such that $\# \Lambda \cap B(y,R) > 0$ for all $y \in \mathbb{R}^d$ ;
\item[ii.] there exists $r>0$ such that $d(y_i,\Lambda\setminus \{y_i\})\geq r$ for all $i \in \mathbb{Z}^d$.
\end{itemize}
In particular, to each admissible set of points $\Lambda$ one can
associate a Delaunay triangulation $\mathcal{D}(\Lambda)$.
A stochastic lattice $\mathcal{L} : \Xi \to
\left(\mathbb{R}^d\right)^{\mathbb{Z}^d}$ is said to be admissible,
if for $P$-almost every $\omega \in \Xi$, $\mathcal{L}(\omega)$
is an admissible set of points, and if the Delaunay triangulation is
regular in the sense of the interpolation theory. 

\medskip

Given a realization $\Lambda(\omega)$ of a stochastic lattice, one may rescale the lattice by a factor $h$, setting $y_i^h(\omega)=hy_i(\omega)$. Let us then define $\Lambda(\omega)^h(\Omega)=\{y_i^h \in \mathbb{R}^3, y_i \in \Lambda(\omega) \mbox{ and }y_i^h \in \Omega \}$, which is the intersection of the rescaled lattice with $\Omega$. 
Using the notations of B\"ol and Reese, one may think of a mesh $\calT_h$ as the Delaunay triangulation of $\Lambda(\omega)^h(\Omega)$.
 
\medskip

In what follows, we assume that $\calT_h$ is related to a stochastic network in the way described above and we add an index $\omega$ to make the stochastic dependence more explicit.
Futhermore we will make some hypotheses related to the stochastic network and the probability space:
\begin{itemize}
\item[(a)] there exists a stationary translation group that acts on the stochastic lattice and which is ergodic for $P$;
\item[(b)] there exists a stationary group of rotations that acts on the stochastic lattice (the stochastic lattice is then said to be rotation invariant on average).
\end{itemize}
Roughly speaking, the translation invariance allows us to obtain a deterministic limit whereas the invariance by rotation implies that the energy density at the limit is isotropic. 
We refer the interested reader to~\cite{Alicandro-Cicalese-Gloria-07b} for the precise formulation of the assumptions and the proofs of the following result.

\medskip

Within hypothesis~(a), there exists an homogeneous quasiconvex energy density $W_{hom}$, which is frame-invariant and satisfies a growth condition of order $p$, such that:
\begin{itemize}
\item For almost all $\omega \in \Xi$, $$\dps{\lim_{h\to 0} \inf \left\{ E_h^\omega (v_h^\omega),v_h^\omega \in V_h^\omega+BC \right\} = \inf \left\{ E_{hom}(v),v \in V+BC \right\}},$$ where $V$ is the Sobolev space $W^{1,p}(\Omega)$, and 
$$E_{hom}(v)=\int_\Omega W_{hom}(\nabla v); $$
\item $W_{hom}$ is given by the asymptotic discrete homogenization formula
$$
\begin{array}{rcl}
W_{hom}(\xi)&=& \lim_{h \to 0} \int_{\Xi} \inf \{ E_h^\omega(v_h^\omega), v_h^\omega \in V_h^\omega, v(x_i)=\xi\cdot x_i \\ 
\\
&&\qquad \qquad \qquad \mbox{ if }d(x_i,\partial \Omega)\leq 2hR \}dP(\omega).
\end{array} 
$$
\end{itemize}
In addition, minimizers $u_h^\omega$ of $E_h^\omega$ on $V_h^\omega+BC$ weakly converge in $W^{1,p}(\Omega)$ to minimizers $u_{hom}$ of $E_{hom}$ on $V+BC$.
This proves in particular the convergence of the solutions of the finite element modelling of rubber introduced by B\"ol and Reese.

\medskip

In addition, if~(b) holds, the stochastic network is rotation invariant on average, and $W_{hom}$ is isotropic.

\section{Conclusion}

In the first section, we have recalled the finite element modelling of rubber developed by B\"ol and Reese and we have pointed out some convergence issues related to their formulation. 
In the following sections, we have analyzed the asymptotic behaviour of the model.
In particular, we have identified two cases for which their model provides a finite element approximation of 
 an underlying continuous model.
The continuous models obtained both satisfy the frame invariance property and the usual incompressibility behaviour used for rubber modelling (up to some technicalities detailed in \cite{Alicandro-Cicalese-Gloria-07b}).
However, only the stochastic model is ensured to yield an isotropic energy density at the limit.
We have also provided a periodic example which is not isotropic at the limit.

% \begin{figure}[tbh]
%   \centering
%   \includegraphics[height=3cm]{Image1.eps}
%   \caption{Figure caption\label{fig1}}
% \end{figure}

\bibliographystyle{unsrt}
\bibliography{book}

\end{document}